\documentclass[12pt]{article}
\usepackage[round]{natbib}
\usepackage{amsmath,amssymb,amsfonts} 
\usepackage{graphicx, amsthm}                 
\usepackage{color}                    
\usepackage{bm}
\usepackage{titletoc}
\usepackage{minitoc,lipsum}
\usepackage{xr}
\usepackage{refcheck}
\usepackage{textcomp, latexsym, parskip,subfig,pdflscape,fancyhdr}
\usepackage{graphicx}

\norefnames

\def\bB{ {\bf B} }
\def\bE{ {\bf E} }
\def\bU{ {\bf U} }
\def\bW{ {\bf W} }
\def\bX{ {\bf X} }
\def\bY{ {\bf Y} }
\def\bb{ {\bf b} }
\def\be{ {\bf e} }
\def\bx{ {\bf x} }

\def\b0{ {\bf 0} }

\def\bOmega{ {\bm \Omega} }

\def\be{ {\bm e} }
\def\bI{ {\bm I} }

\def\bx{ {\bm x} }

\def\by{ {\bm y} }

\begin{document}
\title{\textbf{Simultaneous Confidence Tubes for Comparison of Several Multivariate Linear
Regression Models}}
\author{
J. Peng$^a$, W. Liu$^{b}$, F. Bretz$^{c}$, A. J. Hayter$^d$ \\
$^a$ Department of Mathematics and Statistics\\
Acadia University, Wolfville, NS, B4P 2R6, Canada\\
jianan.peng@acadiau.ca\\
$^{b}$S3RI and School of Mathematics \\
University of Southampton, UK\\
$^{c}$Novartis Pharma AG, Basel, 4002, Switzerland\\
$^d$Dept of Statistics and Operations Technology\\
University of Denver, 80208-8921, USA
}
\date{}
\maketitle

\vfill
\eject

\begin{abstract}
Much of the research on multiple comparison and simultaneous inference in the
past sixty years or so has been for the comparisons of
several population means. Spurrier (1999) seems to be the first
to study the multiple comparison of several simple
linear regression lines by using simultaneous confidence bands.
In this paper, the work of Liu {\it et al.} (2004) for finite
comparisons of several univariate linear regression models by using
simultaneous confidence bands has been extended to finite comparison of
several multivariate linear regression models by using simultaneous
confidence tubes.
We show how simultaneous confidence tubes can be constructed to allow
more informative inferences for the
comparison of several multivariate linear regression models than the
current approach
of  hypotheses testing.
The methodologies are illustrated with examples.
\end{abstract}


\vskip 0.25in

\section{Introduction}\label{section intro}

The bulk of the work on simultaneous inference and multiple comparisons to date is
for comparing the means of $k (\ge 3)$ populations, following the work of
Tukey (1953) on pairwise comparisons of $k$ population means,
of Dunnett (1955) on comparisons of several
means with a control mean,
and of Scheff\'e (1953) on all-contrast comparisons among the population means.
Miller (1981), Hochberg and Tamhane (1987),
Westfall and Young (1993), Hsu (1996) and Bretz {\it et al.} (2011)
are excellent references of the work in this area.
Spurrier (1999) seems to have been the first to work on the simultaneous
comparison of several simple linear regression lines by using a set
of simultaneous confidence bands.
Since then, Spurrier's (1999) work has been
extended in several directions; see, for example, Spurrier (2002), Bhargava and Spurrier (2004), Liu {\it et al.} (2004) and Lu and Kuriki (2017).
In particular, Liu {\it et al.} (2004) use simultaneous confidence bands for
finite comparisons of several univariate linear regression models, which is directly applicable
for pooling batches in drug stability study (cf. Ruberg and Hsu, 1992) among
many other applications. A
review of the related works is given in Liu (2010, Chapters 5 and 6).
The purpose of this paper is to extend the work of Liu {\it et al.} (2004) on univariate
linear regression models to multivariate linear
regression models, which have wide applications (cf. Anderson, 2003,
 and Raykov and Marcoulides, 2008). Much of the recent work on simultaneous confidence bands for linear regression is almost exclusively for univariate
linear regression models. See, for example, Al-Saidy {\it et al.} (2003), Nitcheva {\it et al.} (2005),
Piegorsch {\it et al.} (2005), Deutsch and Piegorsch (2012),
Peng {\it et al.} (2015) and Dette {\it et al.} (2018).
Liu {\it et al.} (2016)
consider simultaneous confidence band for one multivariate linear regression model over the whole covariate
region, and calls the confidence band ``confidence tube'' to reflect its true shape.

Assume the $i$-th multivariate linear regression model, corresponding to the
$i$-th treatment group,  is given by
$$
{\bY}_i = \bX_i {\bB}_i + {\bE}_i, \ \ \ i=1,\cdots , k
$$
where $\bY_i = (\by_{i,1}, \cdots , \by_{i, n_i})^T$ with
$\by_{i,j}^T = (y_{i,j,1}, \cdots, y_{i,j,m})$ being the observations on the $m$ response variables of the $j$-th individual
in the $i$-th treatment group,
$\bX_i$ is a $n_i\times (p+1)$ full column-rank design matrix with the first column given
by $(1, \cdots , 1)^T$ and the $l(\ge 2)$th column given by
$(x_{i,1,l-1}, \cdots, x_{i,n_i,l-1})^T$, $\bB_i =(\bb_{i,1}, \cdots, \bb_{i,m})$
with $\bb_{i,j}^T = (b_{i,j,0},\cdots,
b_{i,j,p})$ being the regression coefficients for the $j$-th response variable in the $i$-th treatment group, and
$\bE_i =(\be_{i,1},\cdots,\be_{i,n_i})^T$ with all
the $\{ \be_{i,j}, j=1,\cdots,n_i, 
 i=1,\cdots,k\}$ being iid multivariate normal $N_m({\bf 0},
\bOmega)$ random vectors. Since $\bX_i^T \bX_i$ is non-singular, the least
squares estimator of $\bB_i$ is given by ${\hat \bB_i}
=(\bX_i^T\bX_i)^{-1}\bX_i^T\bY_i, i=1,\cdots,k$. Let
the pooled estimator of the unknown error-vector covariance matrix $\bOmega$ be
$\hat{\bOmega} = \sum_{i=1}^k (\bY_i - \bX_i \hat{\bB}_i)^T (\bY_i - \bX_i \hat{\bB}_i)/ \nu$
with $\nu = \sum_{i=1}^k
(n_i-p-1)$. Then $vec(\hat{\bB}_i) \sim N( vec(\bB_i), \bOmega \otimes (\bX_i^T\bX_i)^{-1})$
where $vec({\bf A})$ denotes the resultant vector from stacking the columns of the
matrix ${\bf A}$, $\nu \hat{\bOmega}$ has the Wishart distribution ${\bf W}(\bOmega, \nu)$,
and $\hat{\bB}_1, \cdots, \hat{\bB}_k$ and $\hat{\bOmega}$ are independent. All these results,
which are generalization of the univariate regression with $m=1$ response variable to
the multivariate regression with $m (\ge 1)$ response variables,
can be found in the excellent book by Anderson (2003).

Our objective is to construct a set of simultaneous confidence tubes (SCTs) for
$$
\bx^T\bB_i
 - \bx^T\bB_j = (1,x_1,\cdots,x_p)\bB_i
 -  (1,x_1,\cdots,x_p)\bB_j, \ \ \ (i,j)\in \Lambda
$$
over a given covariate range $x_l \in [a_l, b_l], l=1,\cdots,p$, where
$\Lambda$ is an index set that determines the comparison of interest.
For example, if the pairwise comparison is of interest then
$\Lambda = \{ (i,j): 1 \le i \ne j \le k \}$; if the comparison with a control, say,
the second to $k$-th regression models with the first regression
model, is of interest then
$\Lambda = \{ (i,j): 2\le i\le k, j=1 \}$; if the
successive comparison
of the $k$ regression models is of interest then
$\Lambda = \{ (i,i+1): 1\le i\le k-1 \}$.
We construct the following set of SCTs
\begin{eqnarray}
&& \left[ (\bx^T \bB_i - \bx^T \bB_j) - (\bx^T \hat \bB_i - \bx^T \hat \bB_j) \right]
   \left( \nu \hat{\bOmega} \right)^{-1}
   \left[ (\bx^T \bB_i - \bx^T \bB_j) - (\bx^T \hat \bB_i - \bx^T \hat \bB_j) \right]^T \ \ \ \ \ \ \ \ \nonumber \\
&& \le \, c \,\bx^T \left[ (\bX_i^T\bX_i)^{-1} + (\bX_j^T\bX_j)^{-1} \right] \bx, \ \ \forall x_l \in [a_l, b_l]
   \ \, {\rm for}\ l=1, \cdots, p \ \, {\rm and}\ \, \forall (i,j) \in \Lambda \ \ \ \ \ \ \ \
\end{eqnarray}
where $c$ is the critical constant suitably chosen so that the confidence
level of this set of SCTs is equal to
$1-\alpha$.

When there is only $m=1$ response variable, the set of SCTs in (1) becomes the
set of simultaneous confidence bands given
in Liu {\it et al.} (2004).  The textbook approach to the
comparison of $k$ multivariate regression models
(cf. Anderson, 2003, and Raykov and Marcoulides, 2008)
is to perform a hypotheses test
of $H_0: \bB_1 = \cdots = \bB_k$ against the alternative $H_a: \ {\rm not}\ H_0$.
It is shown in this paper that
the SCTs in (1) allow more detailed and informative inferences than the dichotomous inference,
rejection or non-rejection of $H_0$, of a hypotheses test.

The outline of the paper is as follows. Section 2 discusses  the computation
of the critical constant $c$ using simulation. Section 3 focuses on the
comparison of two models (i.e. $k=2$), and illuminates the relationship between Roy's (1953) test
and the SCT in (1) in this case. We provide a real data example
to illustrate the advantages of the SCTs approach over hypotheses testing.
Section 4 considers the comparison of $k(\ge 3)$ models. Again,  we use an example  to
illustrate the versatile and informative inferences the SCTs approach allows.
Finally, Section 5 contains concluding remarks.

\section{ Determination of the critical constant c}

Note that the confidence level of the SCTs in (1) is
given by $P\{ T < c \}$ where
\begin{equation}
T = \sup_{(i,j)\in \Lambda}\ \sup_{x_l \in [a_l,b_l],\, l=1,\cdots,p} T_{i,j}(\bx)
\end{equation}
with
$$ 
T_{i,j}(\bx) = { \left[ (\bx^T \bB_i - \bx^T \bB_j) - (\bx^T {\hat \bB}_i - \bx^T {\hat \bB}_j) \right]
   \left( \nu \hat{\bOmega} \right)^{-1}
   \left[ (\bx^T \bB_i - \bx^T \bB_j) - (\bx^T \hat \bB_i - \bx^T \hat \bB_j) \right]^T
\over
   \bx^T \left[ (\bX_i^T\bX_i)^{-1} + (\bX_j^T\bX_j)^{-1} \right] \bx
}
$$ 

Denote $\bU_i = (\bX_i^T \bX_i)^{1/2}(\hat{\bB}_i - \bB_i) \bOmega^{-1/2}, i=1, \cdots, k$.
Then straightforward manipulation shows that $vec( \bU_i ) \sim N( {\bf 0}, \bI_{(p+1)m})$ and
it is clear that $\bU_1, \cdots, \bU_k$ are independent since $\hat{\bB}_1, \cdots, \hat{\bB}_k$
are independent. Now $T_{i,j}(\bx)$ can be rewritten as
\begin{equation}
{ \bx^T \left[ (\bX_i^T\bX_i)^{-1/2}\bU_i - (\bX_j^T\bX_j)^{-1/2} \bU_j \right]
   \left( \bOmega^{-1/2}\nu \hat{\bOmega} \bOmega^{-1/2} \right)^{-1}
   \left[ \bU_i^T(\bX_i^T\bX_i)^{-1/2} -  \bU_j^T(\bX_j^T\bX_j)^{-1/2} \right] \bx
\over
   \bx^T \left[ (\bX_i^T\bX_i)^{-1} + (\bX_j^T\bX_j)^{-1} \right] \bx
}
\end{equation}
By noting that $\bOmega^{-1/2}\nu \hat{\bOmega} \bOmega^{-1/2} \sim \bW ( \bI_m, \nu )$, the
distributions of $T_{i,j}( \bx )$ and so $T$ do not depend on the unknown parameters
$\bB_1, \cdots, \bB_k$ and $\bOmega$ of the $k$ regression models.

So the critical value $c$ is the $(1-\alpha)$ quantile of $T$ and
can be computed by simulation as the $(1-\alpha)$ sample quantile $\hat{c}$ of a large
simulated sample of $r$ independent replicates $T_1, \cdots, T_r$ of $T$,
using the expressions (2) and (3). It is
known that the sample $(1-\alpha)$ quantile $\hat{c}$ converges to the population $(1-\alpha)$ quantile
$c$ almost surely as $r$ approaches infinity (cf. Serfling, 1980).
This means $\hat{c}$ can be as close to $c$ as required by using a sufficiently large
number $r$ of simulations. For a finite $r$, the accuracy of $\hat{c}$
can be assessed by using the variance of the large sample approximate normal
distribution of $\hat{c}$; see, for example, Liu {\it et al.} (2005) for details.

For the examples given in
this paper, there is only one covariate, $x_1$,  in the
regression models and so both the numerator and
denominator of the $T_{i,j}( \bx )$ in (2) are polynomials of $x_1$. As a result,
$\sup_{x_1 \in [a_1,b_1]} T_{i,j}(\bx)$ in (2)
can be computed very fast by using the
method of Liu {\it et al.} (2008) in each simulation of $T$.
When the regression models have more than one covariate, generic and so less efficient algorithms for finding maximums have to be used and are available in most
numerical software such as R and Matlab.

In the examples in
this paper, $r=1,000,000$ has been used and it takes about 500 seconds on an ordinary Window's PC (Core(TM2) Due CPU P8400@2.26GHz) to compute one $c$. For $r=1,000,000$ and
$\alpha = 0.05$, the standard error (i.e. the square
root of the variance) of $\hat{c}$ is about
0.00004 and so the critical constant is most likely accurate to the fourth decimal place at least.
Alternatively one can use the method of Edwards and Berry (1987) to assess how close the random
variable ${\rm P} \{ T < \hat{c} | \hat{c} \}$ is to $1-\alpha$. This random variable has
a Type I beta distribution with parameters $r-\langle (1-\alpha)r \rangle -1$ and
$\langle (1-\alpha)r \rangle$, and is approximately normal for a large $r$ (Edwards and Berry,
1987). For $r=1,000,000$ and
$\alpha = 0.05$, the standard error of this random variable is $s.e.=0.00022$
and so the true confidence level
by using $\hat{c}$ is almost certainly in the range $1-\alpha \pm 3 s.e. = [0.94934,0.95066] $. All these indicate that
the critical value $\hat{c}$ based on $r=1,000,000$ simulations should be accurate enough for
most applications.

\section{ Comparison of two models}

For comparing $k=2$ models, the set of $1-\alpha$ SCTs in (1) contains just one SCT and is given by
\begin{eqnarray}
&& \left[ (\bx^T \bB_1 - \bx^T \bB_2) - (\bx^T \hat \bB_1 - \bx^T \hat \bB_2) \right]
   \left( \nu \hat{\bOmega} \right)^{-1}
   \left[ (\bx^T \bB_1 - \bx^T \bB_2) - (\bx^T \hat \bB_1 - \bx^T \hat \bB_2) \right]^T \ \ \ \ \ \ \ \ \nonumber \\
&& \le \, c \,\bx^T \left[ (\bX_1^T\bX_1)^{-1} + (\bX_2^T\bX_2)^{-1} \right] \bx, \ \ \forall x_l \in [a_l, b_l]
   \ \, {\rm for}\ l=1, \cdots, p\, .
\end{eqnarray}
This SCT quantifies the magnitude of difference between the two models $\bx^T \bB_1$ and
$\bx^T \bB_2$ over the covariate region  $x_l \in [a_l, b_l]$ for $l=1, \cdots, p$. In particular,
if $\bB_1 = \bB_2 = \bB$ then the zero line $\bx^T (\bB_1 - \bB_2)$ is completely contained in the
SCT with probability $1-\alpha$, which induces the following size $\alpha$ test of $H_0: \bB_1 = \bB_2$
against the alternative
$H_a: {\rm not}\ H_0$: $H_0$ is rejected if and only if the zero line is not
completely contained in the SCT.

For the special covariate region $x_l \in [-\infty, \infty]$ for $l=1, \cdots, p$, i.e. the whole
covariate space $R^p$, the rejection region of $H_0$ of this induced test becomes $L > c$ where
\begin{eqnarray}
&& L =  \sup_{x_l \in R^1,\ l=1,\cdots,p}
\frac{ \left(\bx^T \hat \bB_1 - \bx^T \hat \bB_2\right)
   \left( \nu \hat{\bOmega} \right)^{-1}
   \left(\bx^T \hat \bB_1 - \bx^T \hat \bB_2 \right)^T }{
   \bx^T \left[ (\bX_i^T\bX_i)^{-1} + (\bX_j^T\bX_j)^{-1} \right] \bx
} \nonumber \\
&& = {\rm the\ largest\ eigenvalue\ of}\ \left[ (\bX_i^T\bX_i)^{-1} + (\bX_j^T\bX_j)^{-1} \right]^{-1}
\left( \hat \bB_1 - \hat \bB_2\right)
   \left( \nu \hat{\bOmega} \right)^{-1}
   \left(\hat \bB_1 - \hat \bB_2 \right)^T  \ \ \ \ \ \ \ \  \\
&& = {\rm the\ largest\ eigenvalue\ of}\  \left(\hat \bB_1 - \hat \bB_2 \right)^T
\left[ (\bX_i^T\bX_i)^{-1} + (\bX_j^T\bX_j)^{-1} \right]^{-1}
\left( \hat \bB_1 - \hat \bB_2\right)
   \left( \nu \hat{\bOmega} \right)^{-1}  \ \ \ \ \ \ \ \
\end{eqnarray}
where the equality in (5) follows directly from Mardia {\it et al.} (1979, Theorem A.9.2) and the
equality in (6) follows directly from  Mardia {\it et al.} (1979, Theorem A.6.2).
A few lines of manipulation shows that this test is just Roy's (1953) test
of $H_0: \bB_1 = \bB_2$
against $H_a: {\rm not}\ H_0$; see Anderson (2003, Sections 8.4 and 8.6) for the construction of
Roy's test and other commonly used tests.
This shows that Roy's test is implied by the SCT over the the whole covariate
space. On the other hand, when $a_l = b_l$ for $l=1, \cdots, p$ and so the covariate
region contains just one point, the SCT in (3) becomes the point-wise band. Direct
manipulation, utilizing the generalized $T^2$-statistic (see e.g. Anderson, 2003,
Theorem 5.2.2), shows that the critical constant $c$ in this case is given
by $c = \frac{m}{\nu} f_{m,\nu, 1-\alpha}$ where $f_{m,\nu, 1-\alpha}$
denotes the $1-\alpha$ quantile of the $F$ distribution with $m$ and $\nu$ degrees of freedom.

Of course the magnitude of differences between the two models quantified by the SCT
is more informative than either a rejection or
a non-rejection of $H_0$ of a test. When $H_0$ is rejected,
the SCT allows us to assess over what covariate region the two models are significantly
different and the direction of the difference. Even when $H_0$ is not rejected, which can mean anything but the two models are the same,
the magnitude of difference between the two models derived from the SCT still provides useful information.

Furthermore, it has been argued by numerous statisticians that statistical models often provide good
approximations only over a certain covariate region (cf. Naiman, 1987, and Piegorsch
and Casella, 1988). The SCT in (1)
uses this information in the form of the covariance region
$x_l \in [a_l, b_l]$ for $l=1, \cdots, p$ in its construction.
This SCT is narrower and hence provides sharper
inference over the covariance region of interest
$x_l \in [a_l, b_l]$ for $l=1, \cdots, p$
than the SCT over the whole covariate region $R^p$.

{\bf Example.} Raykov and Marcoulides (2008, pp.192) provide a dataset from a study of how
a set of three intelligence measures, Inductive reasoning (ir),
Figural relations (fr) and Culture-fair tests (cf),
differ across groups at post-test after accounting for the covariate, which is
the pre-test measurement on inductive reasoning. Following Raykov and Marcoulides, a
bivariate linear model of the responses $y_1 = fr$ and $y_2 = cf$ on
the only covariate pre-test measurement on inductive reasoning ($x=ir$) is fitted for each of the
two groups of students: trained and untrained. Note that we do not assume that the two
bivariate linear regression models for the two groups have the same slopes;
otherwise the problem becomes the comparison of the intercepts only
and can be dealt with by the simpler multivariate ANCOVA (cf. Anderson, 2003).
Based on the observations
on 248 students (with 87 students in group one `untrained' and 161 students in group two `trained'),
one can easily compute the estimates
$$
\hat{\bB}_1 =  \left(\begin{array}{cc} 48.404  & 14.689 \\
                                  0.590  & 1.103 \end{array}\right), \ \
\hat{\bB}_2 =  \left(\begin{array}{cc}  53.740  & 22.749 \\
                                  0.637  & 1.040 \end{array}\right), \ \
      \nu \hat{\bOmega}   = \left(\begin{array}{cc} 33075.6 &   20024.8 \\
                                                    20024.8 &   37408.5 \end{array}\right)
$$
with $m=2$, $p=1$ and $\nu = 244$.

The textbook approach for the comparison of the two models for the two groups is to test
$H_0: \bB_1 = \bB_2$ against $H_a: {\rm not}\ H_0$. Specifically, Roy's (1953) test
has its test statistic given by 0.0876, critical value 0.0360
for $\alpha = 0.05$,
p-value 0.0002 and so $H_0$ is rejected. Other commonly used tests (cf. Anderson, 2003, Sections
8.4 and 8.6) also reject $H_0$ with comparable p-values.

\begin{figure}[!htb]
  \caption{\emph{The 95\% SCT: given by the union of all the elliptic discs; the estimate
  $(\bx^T \hat\bB_2 - \bx^T \hat\bB_1)$: given by the straight line in the centre of the SCT;
  the zero line $\bx^T (\bB - \bB)$: given by the other straight line.}}
  \includegraphics[width=1\textwidth]{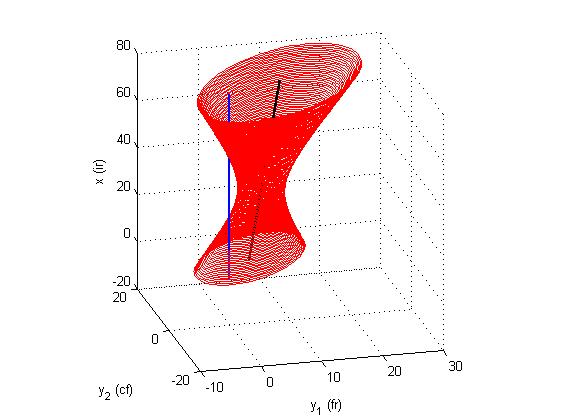}
\end{figure}

The SCT for $(\bx^T \bB_2 - \bx^T \bB_1)$ over the observed covariate range $x \in [0, 78.6]$
in (4) is plotted in Figure 1, with the critical constant $c$ for $\alpha = 0.05$
computed to be 0.0357. 
The SCT is formed by a collection of elliptic discs, one at each $x \in [0, 78.6]$.
The centre of the SCT is given by the straight line $(\bx^T \hat\bB_2 - \bx^T \hat\bB_1)$,
which is the estimate of $(\bx^T \bB_2 - \bx^T \bB_1)$ and also plotted in Figure 1.
Since the zero line $\bx^T (\bB_2 - \bB_1)$ with $\bB_1=\bB_2$, plotted in
Figure 1 by the other straight line,
is not completely contained in the SCT over $x \in [0, 78.6]$, $H_0: \bB_1 = \bB_2$
is also rejected by the SCT. But the SCT provides more information on $(\bx^T \bB_2 - \bx^T \bB_1)$.
For example, by looking at the projection of the SCT to the $(y_1, x)$-plane, plotted in Figure 2,
one can conclude that the trained group (corresponding to $\bx^T \bB_2$) has higher $y_1$ (i.e. fr)
on average than the untrained group (corresponding to $\bx^T \bB_1$) among those students with
the ir score in $[17, 50]$. But the difference between the two groups is not significant
among those students with
the ir score not in the interval $[17, 50]$. Similar observation can be made from the
projection to the $(y_2, x)$-plane. From the SCT, one can also bound
the largest possible difference of $(\bx^T \bB_2 - \bx^T \bB_1)$
over $x \in [0, 78.6]$.

\begin{figure}[!htb]
  \caption{\emph{The projection of the 3-d plot in Figure 1 to the $(y_1, x)$-plane.}}
  \includegraphics[width=1\textwidth]{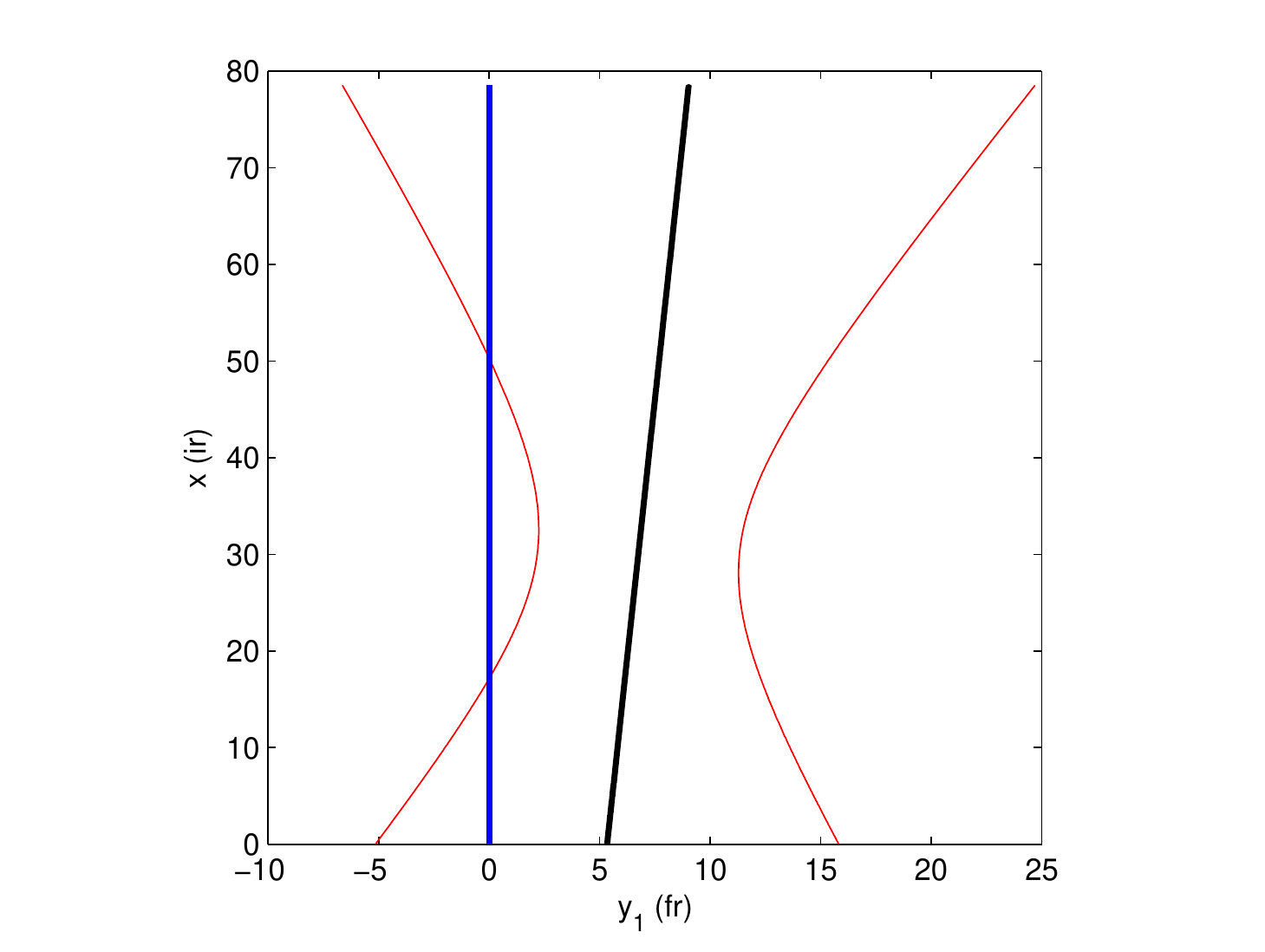}
\end{figure}

It is noteworthy that, in this example, the 95\% SCT over the whole covariate range $x \in [-
\infty, \infty]$ has $c = 0.0360$, which is almost the same as the
$c$ for the SCT over $x \in [0, 78.6]$ given
above. On the other hand, the point-wise band uses $c = \frac{m}{\nu} f_{m,\nu, 1-\alpha}
= 0.0249$, which is $(0.0360-0.0249)/0.0360\% = 31\%$ smaller than the $c=0.0360$. This
indicates the extent to which a SCT over a finite covariate range can potentially be narrower
than the SCT over the whole covariate range.

One can download from
\texttt{http://www.personal.soton.ac.uk/wl/SCTsForMultipComp/}\,
the R codes for the computation of the results and the Matlab codes for
drawing the graphs of this and the next sections.

\section{  Comparison of more than two models}

For comparison of $k (\ge 3)$ models, the SCTs in (1) allow one to assess which models are
different and, if two models are different, over what covariate region and in which direction the
models differ. In comparison, a hypotheses test, such as Roy's test, only concludes whether or not
the $k$ models are different. In this case
there is no clear relationship between the SCTs and Roy's test.
This is not surprising since, even in the simpler situation of  univariate regression,
there is no direct relationship
between simultaneous confidence bands and the usual $F$ test for comparing $k (\ge 3)$ models (cf.
Liu, 2010, Section 6.2).

{\bf Example.}  Continue with the example considered in Section 3. Now
the 161 students
in the `trained' group have actually gone through one of the two different training methods:
the first 80 students were on training method 1 and the other 81 students were on
training method 2. And we are interested in whether the three groups, group 1 -- untrained,
group 2 -- training method 1 and group 3 -- training method 2, are different in terms of
how the responses fr  and cf depend on the covariate
 pre-test measurement ir. Hence we fit
a bivariate linear model of the responses $y_1 = fr$ and $y_2 = cf$  on
the only covariate $x = ir$ to each of the
three groups of students, and we are interested in assessing whether
the three models $\bx^T {\bB}_1$,  $\bx^T {\bB}_2$
and  $\bx^T {\bB}_3$ are the same or not.
Based on the observations
on 248 students (with 87 students in group one, 80 in group 2 and 81 in group 3),
one can easily compute the estimates
\begin{eqnarray}
&& \hat{\bB}_1 =  \left(\begin{array}{cc} 48.404  & 14.689 \\
                                  0.590  & 1.103 \end{array}\right), \ \
\hat{\bB}_2 =  \left(\begin{array}{cc}  51.682  & 23.964 \\
                                  0.681  & 0.995 \end{array}\right), \ \  \nonumber \\
&& \hat{\bB}_3 =  \left(\begin{array}{cc} 56.176  & 21.292 \\
                                  0.582  & 1.092 \end{array}\right), \ \
      \nu \hat{\bOmega}   = \left(\begin{array}{cc} 32908.6 &   20090.3 \\
                                                    20090.3 &   37323.4 \end{array}\right) \nonumber
\end{eqnarray}
with $m=2$, $p=1$ and $\nu = 242$.

\begin{figure}[!htb]
  \caption{\emph{The SCT for $(\bx^T \bB_3 - \bx^T \bB_1)$.}}
  \includegraphics[width=1\textwidth]{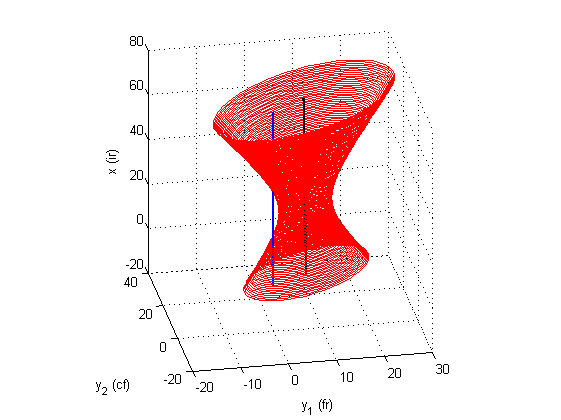}
\end{figure}

The textbook approach for comparing the three models for the three groups is to test
$H_0: \bB_1 = \bB_2 = \bB_3$ against $H_a: {\rm not}\ H_0$.
If Roy's (1953) test is used, then the test statistic is computed to be
0.0899, the critical value for $\alpha = 0.05$ is 0.0536, the
p-value is 0.0017 and so $H_0$ is rejected. But this is all a test can tell us.

To get more information on how the three models differ between themselves, one can use
the SCTs in (1) for pairwise comparison with $\Lambda = \{ (i,j): 1 \le i \ne j \le 3 \}$.
For $\alpha = 0.05$ and the observed covariate region $x \in [0, 78.595]$, the critical
constant $c$ is computed to be 0.0462. So one
can plot the three SCTs for $(\bx^T \bB_2 - \bx^T \bB_1)$, $(\bx^T \bB_3 - \bx^T \bB_1)$ and
$(\bx^T \bB_3 - \bx^T \bB_2)$ over $x  \in [0, 78.6]$, respectively, in order to assess
whether or how any two models differ. For example, Figure 3 plots the SCT for
$(\bx^T \bB_3 - \bx^T \bB_1)$, the straight-line $(\bx^T \hat{\bB}_3 - \bx^T \hat{\bB}_1)$
which is the centre of the SCT, and the zero straight-line
$(\bx^T {\bB} - \bx^T {\bB})$. Since the zero straight-line is not included in the SCT
completely, the two models $\bx^T \bB_3$ and $\bx^T \bB_1$ are significantly different.
By looking at the SCT from different angles, one can observe how the two models differ.
For example, by looking at the projection of the SCT for $(\bx^T \bB_3 - \bx^T \bB_1)$ in the
$(y_1, x)$-plane, given in Figure 4, one can conclude that training method 2 produces
significantly higher fr scores than untrained for the students with $x=ir$
measure in the range
$[20, 41]$. Similarly, by inspecting the SCT for $(\bx^T \bB_2 - \bx^T \bB_1)$, one can
also conclude that training method 1 is also significantly different from untrained
since this SCT does not include the zero line completely. 
However, the SCT for $(\bx^T \bB_3 - \bx^T \bB_2)$ contains the zero line over $x  \in [0, 78.6]$
and so there is no significant difference between the two training methods. Since all
these inferences are based on the three SCTs with a simultaneous confidence level 95\%, one
can claim that all the inferences made are correct simultaneously with confidence level 95\%.

\begin{figure}[!htb]
  \caption{\emph{The projection of the SCT for $(\bx^T \bB_3 - \bx^T \bB_1)$.}}
  \includegraphics[width=1\textwidth]{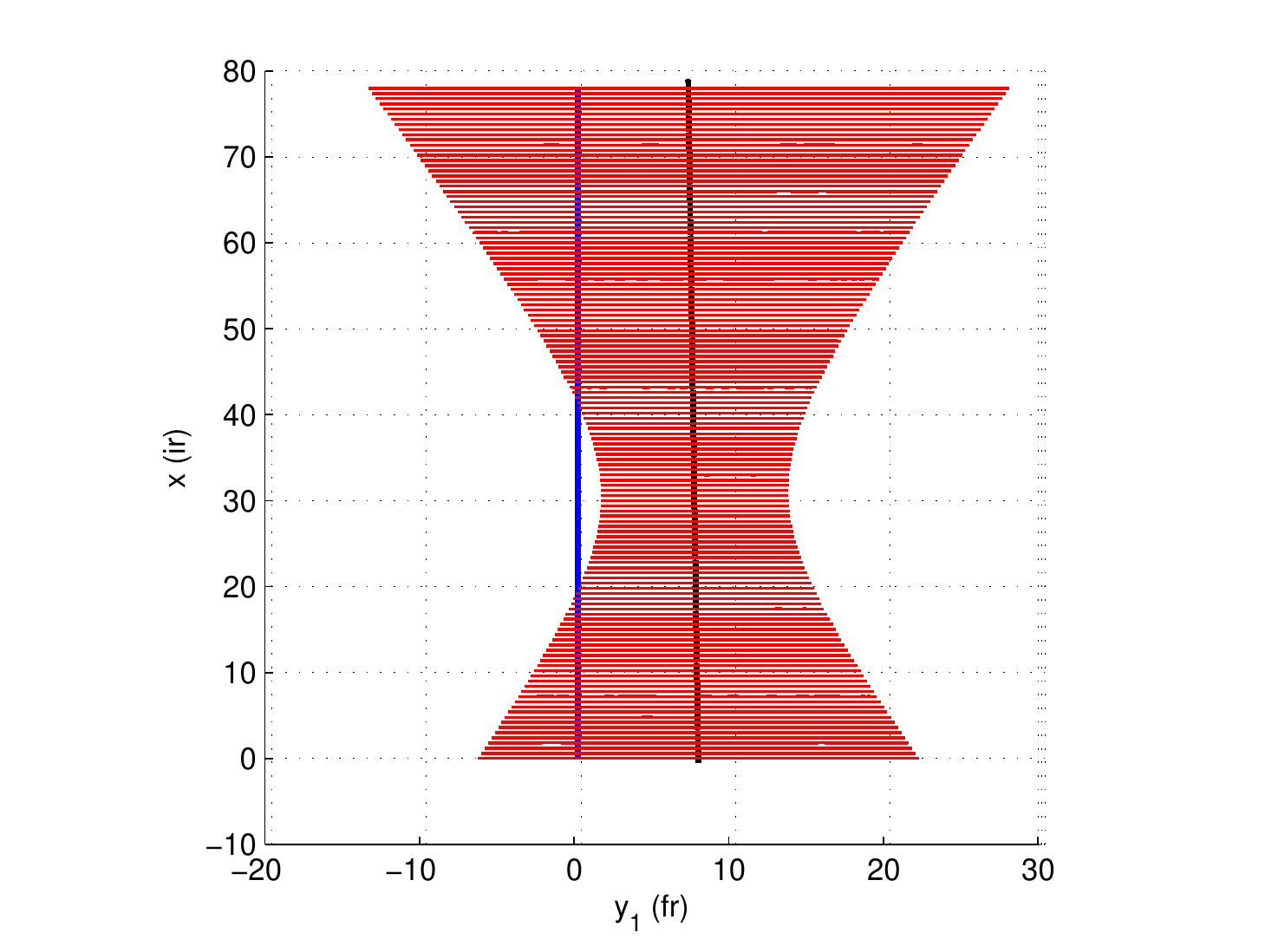}
\end{figure}

Now suppose that one is only interested in whether and how the two training methods are
different from the untrained method. If one uses Roy's test for this purpose, then
the same test, as given above, has to be used with the same conclusion that $H_0$ is rejected.
On the other hand, one can use the SCTs in (1) with $\Lambda = \{ (i,j): i=1, \, 2 \le j \le 3 \}$
specifically for the inferences about $(\bx^T \bB_2 - \bx^T \bB_1)$ and
$(\bx^T \bB_3 - \bx^T \bB_1)$ only.
For $\alpha = 0.05$ and the covariate region $x \in [0, 78.595]$, the critical
constant $c$ is computed to be 0.0424, which is smaller than the
critical constant 0.0462 for pairwise comparisons as expected. Again, one
can look at the two SCTs for $(\bx^T \bB_2 - \bx^T \bB_1)$ and $(\bx^T \bB_3 - \bx^T \bB_1)$ over $x  \in [0, 78.6]$, respectively, to make appropriate inferences. Since in this case one is
interested in two comparisons only, the corresponding SCTs are $(0.0462-0.0424)/0.0424 = 9\%$
narrower and so allow sharper inferences than the SCTs for the three pairwise comparisons given above.
This demonstrates how more informative SCTs can be constructed for particular inferences of interested.

In this particular example, there are two responses $y_1$ and $y_2$ and so all the SCTs can be
plotted in the 3-dimensional space. By inspecting these SCTs directly,
inferences about the comparisons of the
models can be made. When there are more than two responses, the SCTs cannot be plotted in the 3-dimensional
space. This is of course due to the multivariate nature of the problem
as with many other multivariate statistical techniques.
On the other hand, if one is only interested in judge whether the zero line is completely contained
in a SCT, then one can use the multiplicity-adjusted p-values of Westfall and Young (1993) in a way similar
to what is used in Liu (2010, pp.166-168) for the univariate regression case. Specifically,
one first computes the observed value $t_{i,j}$ of
$$
T_{i,j} = \sup_{x_l \in [a_l,b_l], \, l=1, \cdots, p}
\frac{ \left( \bx^T \hat \bB_i - \bx^T \hat \bB_j \right)
       \left( \nu \hat{\bOmega} \right)^{-1}
       \left( \bx^T \hat \bB_i - \bx^T \hat \bB_j \right)^T }{ \bx^T \left[ (\bX_i^T\bX_i)^{-1} + (\bX_j^T\bX_j)^{-1} \right] \bx }
 \, .
$$
One then computes $p_{i,j} = {\rm P}\{ T > t_{i,j} \}$ for $(i,j) \in \Lambda$ by simulating
a large number of replicates of $T$, using the expressions (2) and (3)
as before. Now the SCT for
$(\bx^T \bB_i - \bx^T \bB_j)$ contains the zero line over the given covariate region if and only
if $p_{i,j} > \alpha$. This allows one to judge whether a SCT contains the zero line without looking
at the plot of the SCT.

For example, for the SCTs for pairwise comparisons of the three models, our R program has computed
$p_{1,2}=0.0195$, $p_{1,3}=0.0020$ and $p_{2,3}=0.8420$ (based on $r=1,000,000$ simulations).
From these, one can conclude directly
that the SCTs for $(\bx^T \bB_2 - \bx^T \bB_1)$ and $(\bx^T \bB_3 - \bx^T \bB_1)$ do not
contain the zero line and that the SCT for  $(\bx^T \bB_3 - \bx^T \bB_2)$ does contain the zero line,
which agrees with what one can see from the plots of the SCTs as expected.
For the SCTs for comparisons of the two training methods with the untrained method, our R program has computed
$p_{1,2}=0.0129$ and $p_{1,3}=0.0013$.
From these, one can conclude directly
that the SCTs for $(\bx^T \bB_2 - \bx^T \bB_1)$ and $(\bx^T \bB_3 - \bx^T \bB_1)$ do not
contain the zero line.

\section{ Conclusions}
Much of the research on multiple comparison and simultaneous inference in the
past sixty years or so has been for the comparison of
several population means. Spurrier (1999) studies
the multiple comparison of several simple
linear regression lines by using simultaneous confidence bands.
In this paper, the work of Liu {\it et al.} (2004) for finite
comparisons of several univariate linear regression models by using
simultaneous confidence bands has been extended to finite comparison of
several multivariate linear regression models by using SCTs.
We have demonstrated how the critical constants
for many types of comparison can be easily computed by Monte
Carlo simulation as in Liu {\it et al.} (2004, 2005).

A SCT provides useful information on the difference between
two multivariate linear regression models over a given range of the explanatory
variables. This information can be used to detect differences
between the two models over the range, as illustrated in the examples provided.
Potentially, it can also be used
to establish the maximum difference and hence equivalence of the two models over the range.
A set of SCTs is certainly
more informative than the current textbook approach of a hypotheses test
for comparing several multivariate linear regression models which allows only two
decisions, rejection or non-rejection of $H_0$, and does not take into consideration of
specific comparisons, such as the comparisons of several treatments with one control, that
may be of interest in a problem.

It is also pointed out in this paper that Roy's (1953) test for comparing
two multivariate linear regression models is implied by a  SCT for the difference
of the two models over the whole covariate space. But often comparison over
a finite covariate space is of interest in applications since regression models
are good approximations usually over a finite covariate space only. SCT's utilise
this finite covariate space restriction naturally in its construction.

Many problems warrant further research, for example, the construction of SCTs of
different shapes for different inferential purposes, extending some ideas in the
univariate regression cases (e.g. Liu {\it et al.}, 2009).

\noindent{\bf Acknowledgements:}
The first author's research was partially supported by NSERC of Canada.

\vfill
\noindent{\bf  References}
\vskip 7pt

Al-Saidy, O. M., Piegorsch, W. W., West, R. W. and Nitcheva, D. K. (2003), ‘Confidence bands for low-dose risk
estimation with quantal response data’, {\it Biometrics}, 59, 1056-1062.

Anderson, T.W. (2003). {\it An Introduction to Multivariate Statistical Analysis, 3rd Edition}. Wiley: New York.

Bhargava, P. and Spurrier, J.D. (2004).
Exact confidence bounds for comparing two regression lines with a control regression line on a fixed interval.  {\it Biometrical Journal}, 46, 720-730.

Bretz, F., Hothorn, T. and Westfall, P. (2011). {\it Multiple Comparisons Using R}.
CRC Press: New York.


Dette, H., Mollenhoff, K., Volgushev, S. and Bretz, F. (2018). Equivalence of regression curves.
{\it Journal of the American Statistical Association}, 113, 711-729.

Deutsch, R. and Piegorsch, W. (2012), ‘Benchmark dose profiles for joint action quantal data in quantitative risk assessment’,
{\it Biometrics}, 68, 1313-1322.

Dunnett, C.W. (1955). A Multiple Comparison Procedure for Comparing
Several Treatments With a Control.
{\it Journal of the American Statistical Association}, 50, 1096-1121.

Edwards, D., and Berry, J.J. (1987). The Efficiency of Simulation-based
Multiple Comparisons.
{\it Biometrics}, 43, 913-928.

Hochberg, Y. and Tamhane, A.C. (1987).
{\it Multiple Comparison Procedures}. Wiley: New York.

Hsu, J.C. (1996). {\it Multiple Comparisons: Theory and Methods}.
Chapman $\&$ Hall: New York.

Liu, W. (2010). {\it Simultaneous Inference in Regression }. CRC Press: New York.

Liu, W., Jamshidian, M. and Zhang, Y. (2004).  Multiple comparison of several linear regression models. {\it Journal of the American
Statistical Association},  99, 395-403.

Liu, W., Jamshidian, M., Zhang, Y. and J. Donnelly (2005). Simulation-based simultaneous confidence bands in multiple
 linear regression with predictor variables constrained in intervals. {\it Journal of Computational and Graphical Statistics}, 14, 459-484.

Liu, W., Bretz, F., Hayter, A.J. and Wynn, H.P. (2009).  Assessing non-superiority, non-inferiority or equivalence when comparing two regression models over a restricted covariate region.   {\it Biometrics}, 65, 1279-1287.

Liu, W., Han, Y., Wan, F., Bretz, F. and Hayter, A. J. (2016).
Simultaneous confidence tubes in multivariate linear regression. {\it Scandinavian Journal of Statistics},  43, 879-885



Liu, W., Wynn, H.P. and Hayter, A.J. (2008).  Statistical inferences for linear regression models when the covariates have functional relationships: polynomial regression. {\it  J. of Statistical Computation and Simulation}, 78(4), 315-324.

Lu, X. and Kuriki, S. (2017). Simultaneous confidence bands for contrasts between several nonlinear regression curves.
{\it Journal of multivariate analysis}, 155, 83-104.

Mardia, K.V., Kent, J.T. and Bibby, J.M. (1979). {\it Multivariate Analysis}. Academic Press: New York.

Miller, R.G. (1981).
{\it Simultaneous Statistical Inference}.
Springer-Verlag: New York.

Naiman, D.Q. (1987). Simultaneous confidence-bounds in multiple-
regression using predictor variable constraints. {\it Journal of the American
Statistical Association}, 82, 214-219.

Nitcheva, D. K., Piegorsch, W. W., West, R. W. and Kodell, R. L. (2005), ‘Multiplicity-adjusted inference in risk
assessment: Benchmark analysis with quantal response data’, {\it Biometrics}, 61, 277-286.

Peng, J., Robichaud, M. and Alsubie, A. (2015), ‘Simultaneous confidence bands for lower-dose risk estimation with
quantal data’, {\it Biometrical Journal}, 57, 27-38.

Piegorsch, W., West, R., Pan, W. and Kodell, R. (2005), ‘Low dose risk estimation via simultaneous statistical inferences’,
{\it Journal of the Royal Statistical Society, Series C}, 54, 245-258.

Piegorsch, W.W. and Casella, G. (1988). Confidence bands for logistic
regression with restricted predictor variables. {\it Biometrics}, 44, 739-750.

Raykov, T. and Marcoulides, G.A. (2008). {\it An Introduction to Applied Multivariate Analysis}. Routledge: New York.

Roy, S.N. (1953). On the heuristic method of test construction and its use in multivariate
analysis.
{\it Annals of Mathematical Statistics}, 24, 220-238.

Ruberg, S.J. and Hsu, J.C. (1992). Multiple comparison procedures for pooling batches in stability studies. {\it Technometrics}, 34, 465-472.

Scheff\'e, H. (1953).
A method for judging all contrasts in the
analysis of variance.
{\it Biometrika}, 40, 87-104.

Serfling, R. J. (1980).
{\it Approximation Theorems of Mathematical Statistics}.
Wiley: New York.

Spurrier, J.D. (1999). Exact Confidence Bounds for All
Contrasts of Three or More Regression Lines.
{\it Journal of the American
Statistical Association}, 94, 483-88.

Spurrier, J.D. (2002). Exact multiple comparisons of three or more regression lines: pairwise comparisons and comparisons with a control. {\it Biometrical Journal}, 44, 801-812.

Tukey, J.W. (1953). {\it The Problem of Multiple Comparisons}, Dittoed
manuscript of 396 pages, Department of Statistics, Princeton
University.

Westfall, P.H., and Young, S.S. (1993),
{\it Resampling-Based Multiple Testing: Examples and Methods for
P-Value Adjustment}. Wiley: New York.


\vfill
\eject

\end{document}